\documentclass[12pt]{article}
\usepackage{amssymb,bm,epsf,epsfig,graphicx}
\input epsf.sty
\newcommand{\re}{\mathop{\rm Re}\nolimits}

\def\ket#1{|#1 \rangle}
\def\aver#1{\left\langle\, #1 \,\right\rangle}

\def\half{\frac{1}{2}}

\def \be {\begin{equation}}
\def \ee {\end{equation}}
\def \bea {\begin{eqnarray}}
\def \eea {\end{eqnarray}}
\def \bdm {\begin{displaymath}}
\def \edm {\end{displaymath}}

\def \nn {{\mathbb N}}

\def \rr {{\mathbb R}}

\def \ll {{\cal L}}
\def \dd {{\cal D}}

\def \hh {{\cal H}}

\def \bb {{\cal B}}

\def \oo {{\cal O}}

\def \eee {{\cal E}}

\def \sss {{\cal S}}

\def \sf  {string field }
\def \sft {string field theory }

\begin{document}
\vskip 2.1cm

\centerline{\large \bf Algebraic solutions in Open String Field Theory ---}
\bigskip
\centerline{\large \bf  --- a lightning review}
\vspace*{8.0ex}

\centerline{\large \rm
Martin Schnabl\footnote{Email: {\tt schnabl.martin at gmail.com}}
}

\vspace*{8.0ex}

\centerline {\large \it Institute of Physics AS CR, Na Slovance 2, Prague 8, Czech Republic}
\vspace*{2.0ex}


\vspace*{6.0ex}

\centerline{\bf Abstract}
\bigskip
In this short talk we review basic ideas of string field theory with the emphasis on the recent developments. We show how without too much technicality one can look for analytic solutions to Witten's open string field theory. This is an expanded version of a talk given by the author over the last year at a number of occasions\footnote{Parts of this work have been presented at the Kavli Institute for Theoretical Physics, the Aspen Center for Physics, the Simons Center for Geometry and Physics and the Yukawa Institute for Physics. We thank these institutes for their warm hospitality.} and notably at the conference {\it Selected Topics in Mathematical and Particle Physics} in honor of Prof. Ji\v{r}\'{\i} Niederle's 70th birthday.
 \vfill \eject

\baselineskip=16pt


\section{What is string field theory?}

The traditional rules of first quantized string theory allow one to compute on-shell perturbative amplitudes, but they tell us little about collective phenomena or non-perturbative effects. Two most prominent examples of such are tachyon condensation (a close relative of the Higgs mechanism) and instanton physics.

String field theory is an attempt to turn string theory into some sort of field theory by writing a field theory action for each of the single string modes and combining them together with very particular interactions. Perturbative quantization of this field theory yields all of the perturbative string theory, and one might hope that one day we could get a truly non-perturbative description of the theory.

One of the most interesting applications of string field theory to date has been in studies of the classical backgrounds of string theory. Traditionally, string theories are defined to be in one-to-one correspondence with worldsheet conformal field theories (CFT's). As such they correspond to the choice of infinitely many couplings in two dimensional worldsheet theory. The condition of vanishing beta functions for all of these couplings is equivalent to Einstein or Maxwell like equations for the classical backgrounds. Given two CFT's, the two corresponding string theories look in general entirely different. For CFT's related by exactly marginal deformations, the two theories may bear some resemblance, but for theories related by relevant deformations it is very hard to see how one background can be a solution of a theory formulated around another background. This is in stark contrast to general relativity, where the Einstein-Hilbert action does not depend on any particular background, but it allows for solutions describing very different geometries.

One of the holy grails of string theory research is a manifestly background independent formulation of string theory. String field theory (SFT) goes half-way towards this goal. It gives us a formulation which is background independent in form (not truly in essence) and which posseses a multitude of classical solutions representing different backgrounds. It is defined using the data of a single reference CFT. It is analogous to writing the Einstein-Hilbert action and substituting the metric $g_{\mu\nu}(x)$ with $g_{\mu\nu}^{\mathrm{ref}}(x)+h_{\mu\nu}(x)$. The fundamental reason for this difficulty is that what are the field theoretic degrees of freedom in string theory depends on the background, unlike in general relativity. Following Sen and Zwiebach \cite{BI1,BI2}, it is believed that the space of classical solutions of SFT is in one-to-one correspondence (modulo gauge symmetries and perhaps dualities) with worldsheet CFT's.

In this short talk we review the amazingly simple construction of a class of solutions that can be determined purely algebraically. These are just the first steps in a long program of constructing and classifying all solutions and relating them to some CFT's. In sections \ref{sec_well} and \ref{sec_ex} we add in a little bit of original material. Borrowing a few theorems from the theory of distributions and the Laplace transform we are able to shed novel light on what the space of allowed string fields should look like. This seemingly academic question is actually important for distinguishing gauge trivial and non-trivial solutions.

\section{Pr\'{e}cis of OSFT}

One of the best understood \sf theories is Witten's covariant Chern-Simons type string field theory \cite{Witten} for open bosonic string.\footnote{There are many other string field theories, also for closed strings or superstrings, and some theories have more than one description, often non covariant. Some are also non-polynomial.}

As is well known, quantization of a single classical string is somewhat subtle, due to the reparametrization invariance of the worldsheet action. This gauge symmetry can be fixed in a number of ways. In the covariant quantization procedure one has to gauge fix the worldsheet metric $h_{\alpha\beta}$ and introduce the worldsheet Fadeev-Popov ghost fields $b$ and $c$. The Virasoro constraints $T_{\alpha\beta}=0$ resulting from gauge fixing can then be conveniently imposed using the BRST formalism.\footnote{Alternatively, as in the light cone gauge, one could use the residual infinite dimensional conformal symmetry to gauge fix one of the embedding coordinates and solve the Virasoro constraints algebraically.} The space of physical states of the string is then identified with the cohomology of the BRST operator $Q_B$ acting on the Hilbert space $\hh_{BCFT}$ of the matter-plus-ghost boundary conformal field theory (BCFT) determined by the string background. Interestingly, and not for trivial reasons, this BCFT is the most convenient starting point for \sf theory.

The classical degrees of freedom of open string field theory are fields associated to quantum states of the first quantized open string. It is very convenient to work with the extended space $\hh_{BCFT}$, which contains not only physical states of the string but also various other states. Interestingly, these turn into auxiliary and ghost fields of string field theory. All the fields are neatly assembled into a string field
\be
\ket{\Psi} = \sum_i \int d^{p+1}k \, \phi_i(k) \ket{i,k},
\ee
where the index $i$ runs over all states of the first quantized string in a sector of momentum $k$. The dimensionality of the momentum is $p+1$, as appropriate for open strings ending on a D-$p$-brane. The coefficients $\phi_i(k) $ are momentum space wave functions for particle like excitations of the string, and would become field theory operators if we proceeded to second quantization.

The action of \sft can be written in the form
\be
S = -\frac{1}{g_o^2}\left[ \half \aver{\Psi * Q_B \Psi} + \frac{1}{3} \aver{\Psi*\Psi*\Psi}\right],
\ee
where $g_o$ is the open string coupling constant and $*$ is Witten's star product. The action has enormous gauge symmetry given by
\be
\delta \Psi = Q_B \Lambda + \Psi * \Lambda - \Lambda * \Psi,
\ee
where $\Lambda \in \hh_{BCFT}$ (Grassman even), provided the start product is associative, $Q_B$ acts as a graded derivation and $\aver{.}$ has properties of integration.

To summarize, the basic ingredients that one needs in order to write down Witten's OSFT in a particular background are
\bdm
\hh_{BCFT},\, *,\, Q_B,\, \langle . \rangle.
\edm
For a more comprehensive review, the reader is referred to the excellent reviews \cite{TZ,SenRev}.\footnote{Older reviews are \cite{Thorn,Bonora} and a more recent development appears in \cite{FK}.}

\section{Demystifying the star product}

The star product has always been one of the most difficult ingredients of the string field to understand and to work with. It can be defined very intuitively using the Schr\"odinger presentation of string wave functionals
\be
\left(\Psi_1 \star \Psi_2 \right)\left[X(\sigma)\right] = \int \left[\dd X_{\mathrm{overlap}}\right]\Psi_1\bigl[\hat X(\sigma)\bigr] \Psi_2\left[\check X(\sigma)\right],
\ee
where the hat and check means that the left and right halves of these functions respectively coincide with those of the $X(\sigma)$. It took some years and many research papers to understand exactly whether this path integral makes sense.

\begin{figure}
\begin{center}
\includegraphics{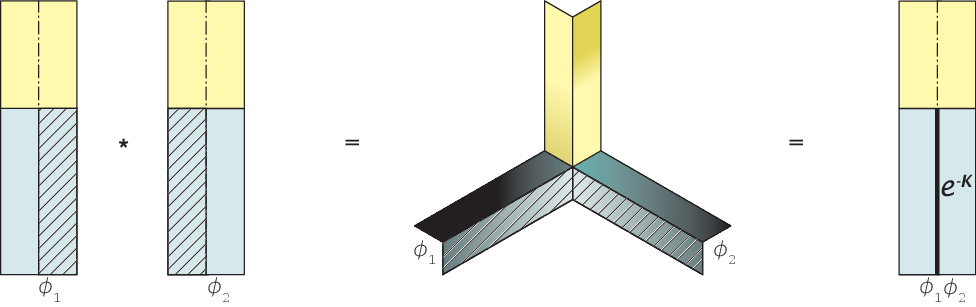}
\end{center}
\caption{\label{fig:sp} Witten's star product is defined by gluing the respective worldsheets.}\end{figure}

There is however a modern definition which makes many of the star product properties manifest.
Let us describe string field theory states as linear combinations of surfaces with vertex operator insertions, such as in Fig. \ref{fig:sp}.\footnote{In order to match with Witten's original definition the $\sigma$ coordinate must run from right to left.} These represent the worldsheets of a single string evolving from the infinite past to the infinite future. A conformal transformation can be used to bring the surface to a canonical form, but this would act nontrivially on the in and out states. We will consider only shapes which have the future (upper) part in the canonical shape of a semi-infinite strip. By putting various vertex operators in the far future and evaluating the path integral over the surface, we can uniquely probe both the shape of the lower part of the surface and what vertex operators are inserted there.

To describe the star product we take two states in the canonical form, cut off the probe strips (in light yellow) and glue the lower parts of the strips along the upper boundaries of the hatched regions. One gets again a state in the form of a surface with insertions, but the shape is different from those we started with. Imagine now factorizing the path integral measure over the worldsheet fields in the hatched area and in the rest of the surface. The path integral over the hatched region is performed first. Then since there are no vertex operators inserted, one can replace its result by an effective term in the worldsheet action, or equivalently as an insertion of some nonlocal line operator. It turns out that this operator can be written as $e^{-K}$, where $K$ is a line integral of the worldsheet energy momentum tensor in some specific coordinates.  The integration extends from the boundary to the midpoint of the worldsheet.

The upshot is that the star multiplication is {\em isomorphic} to operator multiplication. To see this more explicitly, consider two vertex operators $\phi_{1,2}$. The corresponding states $\ket{\phi_{1,2}}$ star multiply as
\be
\ket{\phi_1} * \ket{\phi_2} = \ket{\phi_1 e^{-K} \phi_2}.
\ee
Let us now introduce new (non-local) CFT operators $\hat\phi = e^{K/2} \phi e^{K/2}$ and associated states $\ket{\hat\phi}$. Then clearly
\be
\ket{\hat\phi_1} * \ket{\hat\phi_2} = \ket{\widehat{\phi_1 \phi_2}}.
\ee
Therefore $\phi \to \ket{\hat\phi}$ is the claimed isomorphism.

Usually one does not think of the star product and the operator product as being the same thing.
In particular, there are well known short distance singularities for local vertex operators in nearby points, whereas the star product is usually thought to be much more regular.
Well, thanks to the presence of the $e^{K/2}$ operators, we do indeed get the same type of singularities when we try to star multiply two $\hat\phi$ states as in the vertex operator algebra. The $\hat\phi$ states can be represented by surfaces that differ from those in Fig. \ref{fig:sp} in that the lower bluish part is missing and is replaced by the identification of the left and right parts of the base of the upper semi-infinite strip with a local operator inserted at the midpoint. We say that such string states have no security strips.

\section{Algebraic solutions to OSFT}

To solve the classical equation of motion
\be
Q_B \Psi + \Psi * \Psi =0
\ee
one could try to restrict the huge star algebra to as small subsector as possible. As we first want to study the tachyon condensation, perhaps we should include the vertex operator of the zero momentum tachyon, which is just the $c$-ghost. For the subalgebra to be nontrivial we should also include the non-local operator $K$. One can easily (but not necessarily) add an operator $B$ which is defined by the same type of integral as $K$ with the energy momentum tensor replaced by the $b$-ghost. Together all these elements obey
\bea
&& c^2 =0, \quad B^2=0, \quad \lbrace c, B \rbrace =1 \\
&& \left[ K, B \right] =0, \quad \left[K, c \right] = \partial c.
\eea
The action of the exterior derivative is equally simple
\be
Q_B K =0, \quad Q_B B=K, \quad Q_B c = c K c.
\ee
$Q_B$ is not the only useful derivation. There is also one called  $L^{-}$, which aside of the usual Leibnitz rule also obeys
\be
L^{-} c = -c, \quad L^{-} B = B, \quad L^{-} K = K.
\ee
At a given ghost number, the derivative $L^-$ counts the number of $K$'s and is bounded from below. One could therefore use it to solve the equation of motion order by order in $L^-$ within the subalgebra generated by $K, B, c$.

The simplest possible solution is
\be\label{simplestsol}
\Psi = \alpha c - c K.
\ee
Clearly $Q_B\Psi = \alpha c K c - cK cK = -\Psi^2$. A more general solution has been found by Okawa \cite{Okawa}, following \cite{Schnabl} (see also the works by Erler \cite{Erler1,Erler2})
\be
\Psi = F c \frac{ K B}{1-F^2} c F,
\ee
where $F=F(K)$ is an arbitrary function of $K$. To prove that it obeys the equation of motion requires some straightforward if a bit tedious algebra. The solution can be {\it formally} written in the form
\be
\Psi = (1- F Bc F) Q_B \left(\frac{1}{1-F Bc F}\right),
\ee
which makes the proof of the equations of motion trivial. What is not so trivial is to distinguish a trivial pure gauge solution from the nontrivial solutions. Note that $Bc$ is a projector, in the sense that it squares to itself, and therefore
\be
(1-F Bc F)^{-1} = 1+ \frac{F}{1-F^2} Bc F.
\ee
For a solution to be nontrivial, the factor $F/(1-F^2)$ must be ill defined, whereas the similar looking factor appearing in the string field itself $\tilde F \equiv K/(1-F^2)$ must be well defined.

Before we go in depth into what ill/well defined means, let us discuss another interesting property of these solutions. Expanding \sft around these solutions one obtains a similar looking theory with $Q_B$ replaced by
\be
Q_\Psi = Q_B + \lbrace \Psi, \cdot \rbrace_*.
\ee
The second term acts as a graded star-commutator with $\Psi$. The fluctuations around the vacuum are described by the cohomology of this operator. Interestingly, one could find a homotopy operator $A$ which formally trivializes the cohomology \cite{ES}
\be
A = \frac{1-F^2}{K} B,
\ee
i.e. it obeys $\lbrace Q_\Psi, A \rbrace= 1$. Therefore, formally, any $Q_\Psi$ closed state $\chi$ can be written as $Q_\Psi$ exact: $\chi = Q_\Psi (A \chi)$.

Absence of nontrivial excitations around a given state $\Psi$ is a property expected by Sen's conjectures \cite{SenCon} around the tachyon vacuum, but definitely not around a generic state. We thus find an analogous condition to the one above: $(1-F^2)/K$ should be well defined for the tachyon vacuum, but ill defined for the perturbative vacuum ($F=0$).

Assuming that $F(K)$ is a well defined string field, and adopting a simplifying assumption that $F$ is analytic around the origin, we find that
\be
F(K) = a + b K + \cdots
\ee
gives the tachyon vacuum if $a=1$ and $b \ne 0$, and gives the trivial vacuum for $a \ne 1$. Solutions with $a=1$ and $b=0$ might correspond to something more exotic such us multiple brane solutions, but this has not yet been shown convincingly.

\section{What constitutes a well defined string field ?}
\label{sec_well}

This is still an open question, so we will rather ask a more specific question of when a function $F(K)$ constitutes a well defined string field. Even this question might not have a unique answer, as there are several possible definitions of what might constitute 'good' or 'bad', depending on the context. We define a set of {\it geometric} string field functions $F(K)$ to be those  expressible as\footnote{A much broader class of interesting non-geometric states has been considered recently by Erler \cite{Ted}, inspired by Rastelli \cite{Rastelli}.}
\be\label{F(K)}
F(K) = \int_0^\infty d\alpha f(\alpha) e^{-\alpha K}.
\ee
The name geometric means that we consider superpositions of surfaces, recall that $e^{-\alpha K}$ represents a surface. For $\alpha \in \nn_0$ it is the $\alpha$-th power of the $SL(2,\rr)$ vacuum $\ket{0}$, and for generic $\alpha \ge 0$ one can find a frame (a so called cylinder frame), in which the surface is a strip of size $\alpha$.

What space do we want $f(\alpha)$ to belong to? Obviously a space of functions would be too restrictive, as one would have no hope of representing even the vacuum corresponding to $F(K) = e^{-K}$. The theory of distributions, developed to a large extent by Laurent Schwartz more than sixty years ago, is exactly what we need.

Let us now remind the reader of some of the useful spaces that are introduced in a beautiful treatise \cite{Schw}. Schwartz introduces the following spaces
\bdm
\begin{array}{ccccccccccccccc}
\dd & \subset &  \sss & \subset & \dd_{L^p} & \subset &  \dd_{L^q} & \subset &  \dot\bb  & \subset & \bb & \subset &  \oo_M  & \subset &  \eee \\
\cap && \cap &&\cap && \cap &&\cap && \cap &&\cap && \cap
\\
\eee' & \subset & \oo_c' & \subset &
\dd_{L^p}' & \subset &  \dd_{L^q}' & \subset &  \dot\bb' & \subset &  \bb'& \subset &  \sss' & \subset & \dd'
\end{array}
\edm
where in both lines $ 1 \le p<q<\infty $. The first line denotes spaces of infinitely differentiable functions (in general on $\rr^n$, but here we restrict to $\rr$) which in addition satisfy (together with their derivatives): $\dd$ are of compact support, $\sss$ is the space of fast decaying Schwartz functions, $\dd_{L^p}$ must also belong to $L^p$, $\bb \equiv \dd_{L^\infty}$ are bounded, $\dot\bb$ are both bounded and possess a finite limit at infinity, $\oo_M$ cannot grow too fast, but finally $\eee$ have no restrictions on their growth.

On the second line we have spaces of distributions that are defined as continuous linear functionals on some function space from the first line (in the case of $\oo_c'$ and $\bb'$ with an additional restriction). For example, $\eee'$ is dual to $\eee$ and represents the space of distributions with compact support. $\oo_c'$ are rapidly decaying distributions, i.e. those which together with all their derivatives are bounded even after multiplication with \mbox{$(1+x^2)^{k/2}$} for all $k \in \rr$.  The space $\dd_{L^1}'$ is dual to  $\dot\bb$. $\dd_{L^p}'$ for $p \in (1,\infty)$ are dual to $\dd_{L^{p'}}$ for $p'=p/(p-1)$. A useful characterization of the  $\dd_{L^p}'$ spaces for all $1 \le p \le \infty$ is that they are finite sums of derivatives of functions from $L^p$ (hence $\dd_{L^p} \subset L^p \subset \dd_{L^p}'$), or, equivalently, that their convolution with any $a \in \dd$ belongs to $L^p$. The space  $\dot\bb'$ is dual to $\dd_{L^1}$ and the distributions are characterized by convergence towards infinity.  Whereas $\dd$ is dense in $\dot\bb'$, it is not dense in $\bb' \equiv \dd_{L^\infty}'$, the space of bounded distributions. $\sss'$ is the well known Schwartz space of tempered distributions (those with at most polynomial growth) dual to $\sss$, and finally $\dd'$  is the biggest space of all distributions dual to $\dd$.

After this little expos\'{e}, we are ready to answer the question which space should $f(\alpha)$ in (\ref{F(K)}) belong to. Let us define {\it geometric states} as those for which $f$ is a Laplace transformable distribution, i.e. $f \in \dd'$, but such that there exists $\xi$, for which $e^{-\xi \alpha} f \in \sss'$. We could perhaps have been more generous and have kept only the $f \in \dd'$ condition, but such states would be even more meaningless from the string field perspective. What we need are not the most general geometric states, but more restricted ones.

Let us define a space of {\it $L_0$-safe geometric} string field functions $F(K)$ to be those for which $f \in \dd_{L_1}'$. This conditions comes from considering states $F(K) \ket{I}$ expanded in the Virasoro basis. The coefficients are given by sums of integrals of the form \mbox{$\int_0^\infty d\alpha \, f(\alpha)(\alpha+1)^{-n}$} for $n=0,2,4,\ldots$. For these integrals to be absolutely convergent, we must have $f \in \dd_{L_1}'$. This definition gives us a very nice surprise. Since the star product of string fields $F_1(K)$ and $F_2(K)$ is just a multiplication, in terms of its inverse Laplace transforms $f_1$ and $f_2$ it is a convolution $f_1 * f_2 (x) = \int_0^x dy f_1(y) f_2(x-y)$ (defined in a more sophisticated way when $f_i$ are both actual distributions). Now it is known that the space $\dd_{L_1}'$ is closed under convolution. Therefore the space of $L_0$-safe geometric string fields is closed under star multiplication!

We now proceed to define a space of  {\it $\ll_0$-safe geometric} string field functions $F(K)$ by the condition $f \in \oo_c'$. This condition comes from considering states $F(K) \ket{I}$ expanded in the basis of $\ll_0$ eigenstates (see \cite{Schnabl} for definition), or equivalently from expanding $F(K)$ in the $L^-$ eigenstates (see \cite{ORZ}). We demand that $\int_0^\infty d\alpha \, f(\alpha)\alpha^n$ for $n \in \nn_0$ be absolutely convergent. This forces $f \in \oo_c'$.
Again this space is closed under convolution and hence the space of $\ll_0$-safe geometric string fields is also closed under star multiplication!

Both definitions of safe string fields can be recast in terms of the properties of the function $F(z) = \int_0^\infty f(\alpha) e^{-\alpha z}$. String field function $F(K)$ is

\begin{enumerate}
\item
geometric if and only if there exists $\xi$ such that for all $z$ with $\re z >\xi$, $F(z)$ is holomorphic and $|F(z)|$ is majorized (i.e. bounded) by a polynomial in $|z|$.
\item
$L_0$-safe geometric if and only if $F(z)$ is holomorphic for all $z$ with $\re z > 0$ and  $|F(z)|$ is majorized by a polynomial in $|z|$ for all $\re z \ge 0$.
\item
$\ll_0$-safe geometric if and only if $F(z)$ is holomorphic for all $z$ with $\re z > 0$, $|F(z)|$ is majorized by a polynomial in $|z|$ for all $\re z \ge 0$, and $F(z)$ can be extended to a $C^\infty$ function on the complex half-plane $\re z \ge 0$.
\end{enumerate}
The proof of the first statement can be found in textbooks, and the latter two can be established by a slight modification.

To end this mathematical discussion, let us now give a few examples. The string fields $(1+K)^p$ are both $L_0$ and $\ll_0$-safe geometric since the function $(1+z)^p$ is holomorphic for $\re z > -1$ and obeys all the above conditions. The inverse Laplace transform $f \in \oo_c'$ can be easily computed:
\bea
\frac{1}{\Gamma(-p)} \alpha^{-p-1} e^{-\alpha}, \qquad && p<0 \nonumber \\
\left(1+\frac{d}{d\alpha}\right)^{[p]+1} \left[\frac{1}{\Gamma([p]+1
    -p)} \alpha^{[p]-p} e^{-\alpha} \right], \qquad &&  p>0, \, p \notin \nn \\
\left(1+\frac{d}{d\alpha}\right)^p \delta(\alpha), \qquad && p \in \nn_0. \nonumber
\eea
Here $[p]$ denotes the integer part of $p$, but in fact any integer greater than that can be taken. For $p \in \nn_0$ the inverse Laplace transform actually belongs to the smaller space $\eee'$ of distributions with compact support. Note that for $p>0,\, p \notin \nn$ distribution theory takes care beautifully of the singularities that would be present if one thought of the inverse Laplace transform as a function. Had we considered functions $(1+\gamma^{-1} K)^p$, with $\gamma \in \rr_+$ the domain of holomorphicity would change, the maximal half-plane being $\re z > -\gamma$. The inverse Laplace transform for these functions is  $\gamma f(\gamma \alpha)$. The closer $\gamma$ is to zero, the slower falloff of $f$ we get. If $\gamma$ were taken negative, the inverse Laplace transform would grow exponentially (definitely not what we want in OSFT) which would manifest itself as singularities of $F(z)$ for $\re z >0$.

Another example is the string field $1/\sqrt{1+K^2}$. It is geometric and $L_0$-finite but neither $L_0$ nor $\ll_0$-safe. The inverse Laplace transform is the Bessel function $J_0(\alpha)$. It belongs to the space $\dd_{L^p}'$ for $p>2$. The reason for $L_0$ finiteness is the cancelations due to the oscillatory behavior of the Bessel function. Finally, let us consider string field $\sqrt{1+K^2}$. The inverse Laplace transform is $\delta'(\alpha) + \half\left(J_0(\alpha)+J_2(\alpha)\right)$ and belongs to $\dd_{L^1}'$ but not to $\oo_c'$. Correspondingly it is $L_0$-safe, but not $\ll_0$-safe.

\section{Examples}
\label{sec_ex}

Let us go through some of the simplest examples of OSFT algebraic solutions of the form
\bdm
\Psi = F c \frac{ K B}{1-F^2} c F
\edm
in more detail, and let us try to see what the generalities of the previous section tell us.
Let us remind the reader of the definition $\tilde F \equiv K/(1-F^2)$ in terms of which $\Psi = F c B \tilde F c F$ and the homotopy operator is $A= \tilde F^{-1} B$.
\begin{itemize}
\item $F(K)=a, \quad a \ne 1$

The solution can be simplified as $\Psi = \frac{a^2}{1-a^2} Q(Bc)$.
For this object both $F/(1-F^2)=a/(1-a^2)$ and $\tilde F = \frac{K}{1-a^2}$ are regular, {\it the solution is therefore a pure gauge and is thus the perturbative vacuum}. The would be homotopy operator $A=(1-a^2) B/K$ is singular. This is so, because the inverse Laplace transform of $1/z$ is $1$ (when restricted to $\rr_+$) which does not belong to neither $\oo_c'$, nor $\dd_{L^1}'$.

\item $F(K) = \sqrt{1-\beta K}, \quad \beta \ne 0$

The solution is geometric only for $\beta<0$, but formally for all values one obtains $\Psi = \sqrt{1-\beta K} \beta^{-1} c \sqrt{1-\beta K}$. This is nothing but a real form of the solution (\ref{simplestsol}) with the identification $\beta = \alpha^{-1}$. For this solution both $\tilde F = \beta^{-1}$ and the homotopy operator $A = \beta B$ are very simple and belong to our $L_0$ and $\ll_0$-safe spaces. Thanks to the vanishing cohomology around the vacuum, it {\it is believed to represent the tachyon vacuum} (it can also be shown to be formally gauge equivalent to it) but we have not yet succeeded in computing its energy. The reason for the difficulty is that the string field is too identity like and gives rise to divergences in the energy correlator. Perhaps rephrasing the problem in terms of distribution theory could solve this issue.

\item $F(K) = e^{-K/2}$

The solution in this case is the {\it first discovered analytic solution for the tachyon vacuum} \cite{Schnabl}. There is however one subtlety with this solution. Since
\be
\tilde F = \frac{K}{1-e^{-K}}= \int_0^\infty d\alpha \left(\sum_{n=0}^\infty \delta'(\alpha-n)\right) e^{-\alpha K},
\ee
the inverse Laplace transform of $\tilde F$ does not vanish for large $\alpha$. Consequently $\tilde f \in \bb'$ and does not belong to either $\dd_{L^1}'$ or $\oo_c'$. It is therefore an example of a geometric string field that is neither $L_0$ nor $\ll_0$-safe, but is nevertheless $L_0$-finite. This is also manifested by the fact that $F(z)$ has poles on the imaginary axes. There are interesting consequences to this. Truncating the sum $\sum K e^{-n K}$ at some finite value of $n=N$ one gets a remnant $\frac{K e^{-(N+1)K}}{1-e^{-K}}$ which still contributes significantly to certain observables, in particular to the energy. This is the origin of the so called phantom term in the tachyon vacuum solution. The homotopy operator, on the other hand, is very well defined, and is both $L_0$ and $\ll_0$-safe.

\item $F(K) = \frac{1}{\sqrt{1+K}}$

This is the {\it tachyon vacuum solution} found by Ted Erler and the author \cite{ErS}
\be
\Psi = \frac{1}{\sqrt{1+K}} c B (1+K) c \frac{1}{\sqrt{1+K}}\, .
\ee
The homotopy operator is simply $A=B/(1+K)$, which is perfectly regular. The inverse Laplace transform of $\tilde F = 1+K$ is $\tilde f = \delta(\alpha) + \delta'(\alpha)$, which actually belongs to $\eee'$ and we thus see no need for the phantom term. In fact the energy for this solution can be computed very easily
\bea
E &=& -S =\frac{1}{6} \aver{\Psi, Q \Psi} \\
&=& \frac{1}{6} \aver{(c+Q(Bc))\frac{1}{1+K} c\partial c \frac{1}{1+K}} \nonumber\\
&=& \frac{1}{6} \int_0^\infty dt_1 \int_0^\infty dt_2 e^{-t_1-t_2} \aver{c \, e^{-t_1 K} \, c\partial c \, e^{-t_2 K}} \nonumber\\
&=& -\frac{1}{6\pi^2} \int_0^\infty du u^3 e^{-u} \int_0^1 dv \sin^2 \pi v = -\frac{1}{2\pi^2},
\eea
which is the correct value, minus the tension of the D-brane, according to Sen's conjecture \cite{SenCon}.
The last correlator that we had to evaluate is indeed very simple: two ghost insertions of $c$ and $c\partial c$ on a boundary of a semi-infinite cylinder of circumference $t_1+t_2$ separated by the distance of $t_1$.

\end{itemize}

\section*{Acknowledgments}

\noindent

I would like to thank Ted Erler for collaborating on many of the topics discussed in this review, and for his insightful comments. I would also like to thank the Kavli Institute for Theoretical Physics, the Aspen Center for Physics, the Simons Center for Geometry and Physics, the Yukawa Institute for Physics and APCTP in Pohang, Korea, for their hospitality while I was working on parts of the present work. This research was supported by the EURYI grant GACR EYI/07/E010 from EUROHORC and ESF.


\begin{thebibliography}{99}

\bibitem{BI1}
  A.~Sen,
  ``ON THE BACKGROUND INDEPENDENCE OF STRING FIELD THEORY,''
  Nucl.\ Phys.\  B {\bf 345} (1990) 551.

\bibitem{BI2}
  A.~Sen and B.~Zwiebach,
  ``A Proof of local background independence of classical closed string field
  theory,''
  Nucl.\ Phys.\  B {\bf 414} (1994) 649
  [arXiv:hep-th/9307088].

\bibitem{Witten}
  E.~Witten,
  ``Noncommutative Geometry And String Field Theory,''
  Nucl.\ Phys.\  B {\bf 268}, 253 (1986).

\bibitem{TZ}
  W.~Taylor and B.~Zwiebach,
  ``D-branes, tachyons, and string field theory,''
  arXiv:hep-th/0311017.

\bibitem{SenRev}
  A.~Sen,
  ``Tachyon dynamics in open string theory,''
  arXiv:hep-th/0410103.

\bibitem{Thorn}
  C.~B.~Thorn,
  ``String Field Theory,''
  Phys.\ Rept.\  {\bf 175}, 1 (1989).

\bibitem{Bonora}
  L.~Bonora, C.~Maccaferri, D.~Mamone and M.~Salizzoni,
  ``Topics in string field theory,''
  arXiv:hep-th/0304270.

\bibitem{FK}
  E.~Fuchs and M.~Kroyter,
  ``Analytical Solutions of Open String Field Theory,''
  arXiv:0807.4722 [hep-th].

\bibitem{Okawa}
  Y.~Okawa,
  ``Comments on Schnabl's analytic solution for tachyon condensation in
  Witten's open string field theory,''
  JHEP {\bf 0604} (2006) 055
  [arXiv:hep-th/0603159].

\bibitem{Schnabl}
  M.~Schnabl,
  ``Analytic solution for tachyon condensation in open string field theory,''
  Adv.\ Theor.\ Math.\ Phys.\  {\bf 10} (2006) 433
  [arXiv:hep-th/0511286].

\bibitem{Erler1}
  T.~Erler,
  ``Split string formalism and the closed string vacuum,''
  JHEP {\bf 0705} (2007) 083
  [arXiv:hep-th/0611200].

\bibitem{Erler2}
  T.~Erler,
  ``Split string formalism and the closed string vacuum. II,''
  JHEP {\bf 0705} (2007) 084
  [arXiv:hep-th/0612050].

\bibitem{ES}
  I.~Ellwood and M.~Schnabl,
  ``Proof of vanishing cohomology at the tachyon vacuum,''
  JHEP {\bf 0702} (2007) 096
  [arXiv:hep-th/0606142].

\bibitem{SenCon}
  A.~Sen,
  ``Universality of the tachyon potential,''
  JHEP {\bf 9912} (1999) 027
  [arXiv:hep-th/9911116].


\bibitem{Schw}
L.~Schwartz,
``Th\'{e}orie des distributions'', Hermann, Paris 1966

\bibitem{Ted}
T.~Erler,
{\it to appear}

\bibitem{Rastelli}
L.~Rastelli,
``Comments on the Open String C* Algebra'', Talk at Simons Center for Geometry and Physics Workshop on String Field Theory, Stony Brook, March 23-27, 2009

\bibitem{ORZ}
  Y.~Okawa, L.~Rastelli and B.~Zwiebach,
  ``Analytic solutions for tachyon condensation with general projectors,''
  arXiv:hep-th/0611110.

\bibitem{ErS}
  T.~Erler and M.~Schnabl,
  ``A Simple Analytic Solution for Tachyon Condensation,''
  JHEP {\bf 0910} (2009) 066
  [arXiv:0906.0979 [hep-th]].


\end{thebibliography}
\end{document}